\documentclass[3p,twocolumn]{elsarticle}

\usepackage{lineno,hyperref}
\usepackage{amsmath}
\usepackage{bm}
\usepackage{cuted,multirow}

\modulolinenumbers[5]

\journal{Fundamental Plasma Physics}

\begin{document}

\begin{frontmatter}

\title{Identification of the melting line in the two-dimensional complex plasmas using an unsupervised machine learning method}


\author[mymainaddress]{Hu-Sheng Li}
\author[mymainaddress]{He Huang}
\author[mymainaddress]{Wei Yang}
\author[mymainaddress,mysecondaryaddress]{Cheng-Ran Du\corref{mycorrespondingauthor}}
\cortext[mycorrespondingauthor]{Corresponding author}
\ead{chengran.du@dhu.edu.cn}

\address[mymainaddress]{College of Science, Donghua University, 201620 Shanghai, PR China}
\address[mysecondaryaddress]{Member of Magnetic Confinement Fusion Research Centre, Ministry of Education, 201620 Shanghai, PR China}

\begin{abstract}
Machine learning methods have been widely used in the investigations of the complex plasmas. In this paper, we demonstrate that the unsupervised convolutional neural network can be applied to obtain the melting line in the two-dimensional complex plasmas based on the Langevin dynamics simulation results. The training samples do not need to be labeled. The resulting melting line coincides with those obtained by the analysis of hexatic order parameter and supervised machine learning method. 
\end{abstract}


\end{frontmatter}

\section{Introduction}
A complex plasma is composed of an weakly-ionized gas and micron-sized solid particles~\cite{Merlino:2004,Bouchoule:Book,Melzer:Book}. Due to the higher thermal velocity of electrons, the particles are negatively charged and interact with each other via screening Coulomb (Yukawa) interaction~\cite{Goree:1994,Konopka:2000,Fortov:2005,Shukla:Book}. In the laboratory, monodisperse micorparticles can be levitated in the sheath and confined in a single layer, where gravity force is balanced by the electrostatic force~\cite{Morfill:2009,Ivlev:book}. Under certain conditions, particles can self-organize in a triangular lattice with hexagonal symmetry, forming a two-dimensional (2D) plasma crystal~\cite{Thomas:1996,I:1996}.

Upon heating, a plasma crystal melts and the regular structure vanishes \cite{Melzer:1996,Feng:2010,Couedel:2018,Jaiswal:2019}. In fact, the thermodynamic status of a complex plasma depends on the coupling parameter $\Gamma=Q^2 / 4 \pi \epsilon_0 \Delta k_b T$ and the screening parameter $\kappa=\Delta / \lambda_D$, where $Q$ is the charge, $T$ is the kinetic temperature, $\Delta$ is the interparticle distance, and $\lambda_D$ is the Debye length. The melting line in the phase diagram of the complex plasma has been extensively studied in the past years \cite{Klumov:2010,Castello:2021}. Molecular dynamics simulations have been applied to study the phase diagram in the 2D~\cite{Hartmann:2005} and three-dimensional (3D) complex plasmas~\cite{Hamaguchi:1997}, where the liquid-solid transition was identified by the measurement of the free energy and order parameter, respectively. 

Recently, machine learning methods have been widely applied in the investigations of the complex plasmas~\cite{Dietz:2017,Ding:2021,Himpel:2021}. Particularly, it was employed to obtain the phase diagram in the 2D complex plasmas based on the simulation results and also applied to study the melting in the experiments~\cite{Huang:2022}. The convolutional neural network was applied to the synthesized images of the particle suspension, whose thermodynamics status was labeled for the liquids at high temperatures and for the crystals at low temperatures. Such method is known as supervised learning, which requires prerequisite knowledge of the particle structure for different statuses.

In this paper, we applied an unsupervised machine learning method to obtain the melting line in the phase diagram of the 2D complex plasma. The evolution of the particle positions upon heating were obtained using the Langevin dynamics simulations. The convolutional neural network was applied in an unconventional manner, resulting in the melting temperature at different $\kappa$.

\section{Methods}
A standard Langevin dynamics simulation was employed to simulate the melting process in the 2D complex plasma. The dynamics of individual particles in the suspension were govern by the equation of motion
\begin{equation}
	\label{eq:motion}
	m\ddot{\bm{r}}_{i} +m \nu \dot{\bm{r}}_{i}=-\sum_{j\ne i}\bigtriangledown\phi_{ij} +\bm{L}_{i},         
\end{equation}
where $\bm{r}_{i}$ is the position of particle $i$, $m$ is the particle mass, and $\nu$ is the damping rate, which results from the neutral gas. The Brownian motion was included in the equation and the corresponding Langevin force $\bm{L}_{i}$ at a certain kinetic temperature $T$ was defined by $\left \langle  \bm{L}_{i} \right \rangle=0$ and $\left \langle \bm{ L}_{i}(t)\bm{L}_{j}(t+\tau) \right \rangle = 2\nu m k_b T \delta_{i,j}\delta(\tau)\bm{I} $, where $\delta_{ij}$ is Kronecker delta, $\delta(\tau)$ is the delta function, and $\bm{I}$ is the unit matrix. For the monodisperse particles immersed in the plasma, the interaction between charged particles reads
\begin{equation}
	\label{eq:yukawa}
	\phi_{ij} =  \dfrac{Q^2}{4\pi\epsilon_{0}r_{ij}}\exp(-\dfrac{r_{ij}}{\lambda_D} ) ,
\end{equation}
where $r_{ij}$ is the distance between particle $i$ and $j$ and the constant charge is assumed as $Q=8000$e. In total $6400$ particles were included in the simulation, where the periodic boundary conditions were used. 

\begin{figure}[!ht]
	\includegraphics[width=17pc]{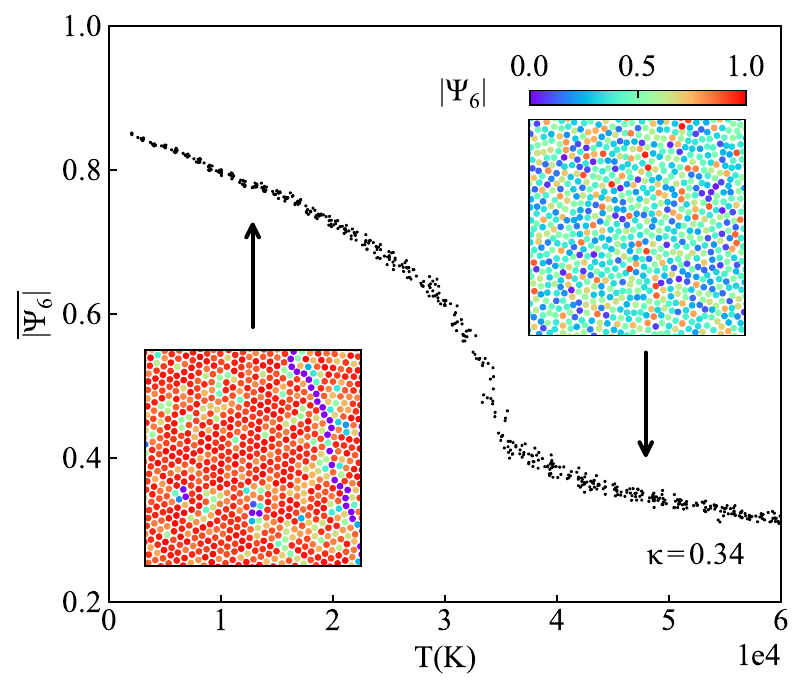}
	\caption{ Dependence of the averaged hexatic order parameter $\overline{|\Psi_6|}$ on the kinetic temperature at $\kappa=0.34$. The structure of the particle suspension for the low (left) and high (right) temperature are demonstrated in the insets, where the hexatic order parameter for the individual particles are colored. In the plasma crystal, the majority of the particles self-organized in the triangular lattice with hexagonal symmetry, where a defect chain was embedded. In the liquid complex plasma, the regular structure was absent and the hexatic order parameters were significantly smaller than unity.   }
	\label{fig:snapshots}
\end{figure}

In the melting process, the temperature rose from $100$~K to $70000$~K. The local structure around particle $i$ could be quantified by the hexatic order parameter
\begin{equation}	
	\label{eq:order}
	\Psi_{6,i}  = \frac{1}{6}\sum_{k=1}^{6}{e^{j6\theta_{k}}}, 
\end{equation}
where six nearest neighboring particles were considered and $\theta_{k}$ is the angle between $\bm{r}_{k}-\bm{r}_{i}$ and the $x$ axis. If particle $i$ is located in the center of  a perfect hexagon cell, its $\Psi_6$ is unity.  As shown in Fig.~\ref{fig:snapshots}, the averaged hexatic order parameter $\overline{|\Psi_6|}$ gradually decreased with the temperature for $T<30000$~K at $\kappa=0.34$. The majority of the particles self-organized in a triangular lattice, where $\Psi_6 \approx 1$. A few defects were embedded in the plasma crystal. The drop of $\overline{|\Psi_6|}$ became faster in the temperature range from $30000$ to $35000$~K, in which the melting transition should happen. As the temperature further increased, $\overline{|\Psi_6|}$ decreased even more slowly than the initial stage of the heating, where the local ordered structure vanished completely, as shown in the inset of Fig.~\ref{fig:snapshots}.

\begin{figure}[!ht]
	\includegraphics[width=18pc]{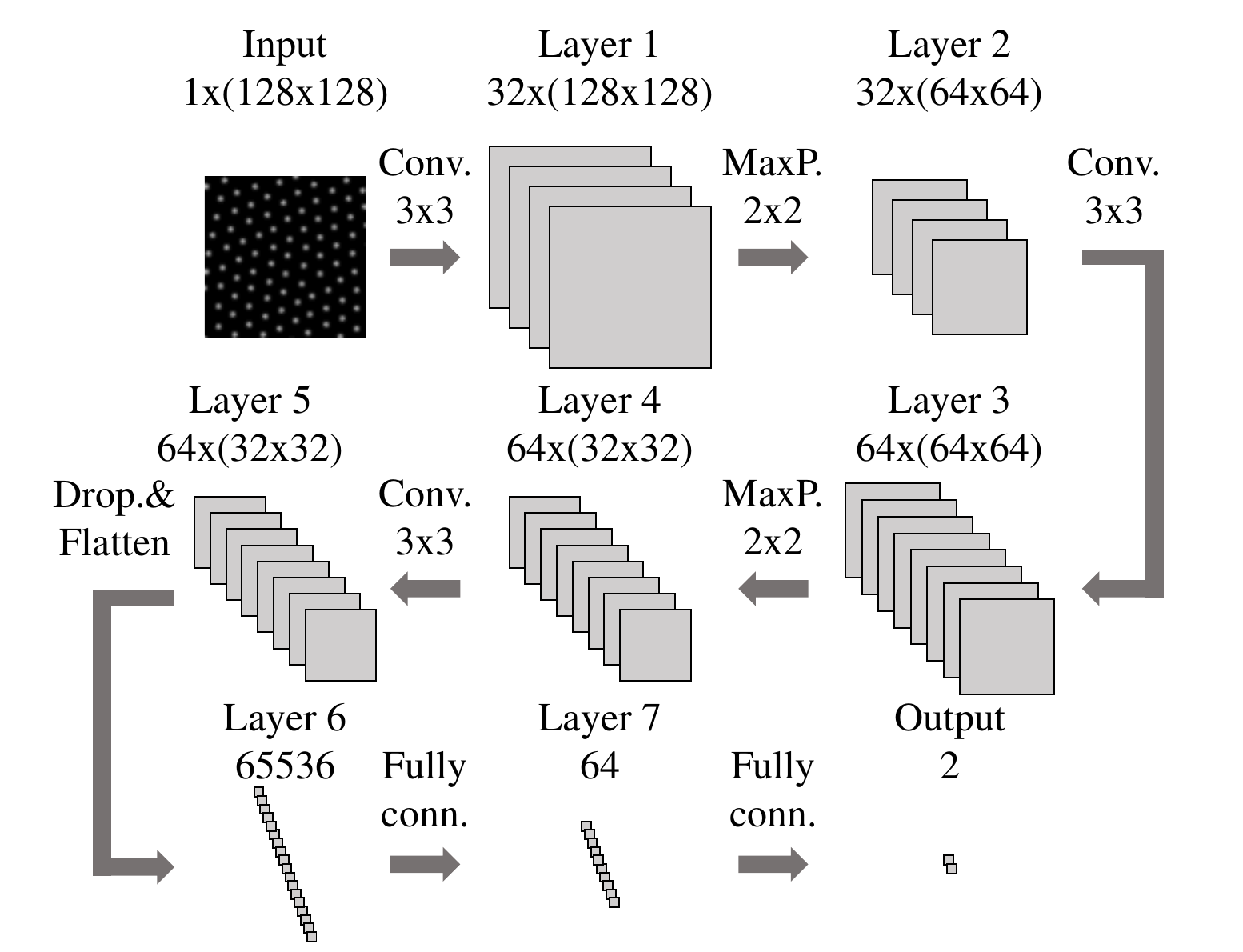}
	\caption{Architecture of the convolutional neural network for the identification of the melting line. Three $3\times3$ kernel 2D convolutional layers (Conv.) were included, followed by max-pooling (MaxP.). Gaussian dropout (Drop) was applied to prevent overfit and the layers were flattened (Flatten). The last two layers were fully connected, leading to a binary classification.}
	\label{fig:achitecture}
\end{figure}

A convolutional neural network (CNN) was applied to train on the simulation results. As CNN is mainly used in the image analysis such as image recognition, object detection, and segmentation~\cite{Krizhevsky:2012}, we converted particle positions with time into sequences of images, which were similar to the experiment recordings~\cite{Feng:2007,Ivanov:2007}. The resulting images had a gray scale, where particles appeared as white spots and the background was black. An example is demonstrated in Fig.~\ref{fig:achitecture}. The details of the image synthesis can be found in Ref.~\cite{Huang:2022}.

The CNN method used in this study contained three $3\times3$ kernel 2D convolutional layers, as shown in Fig.\ref{fig:achitecture}. The rectified linear unit (RELU) was used as the activation function. Max pooling was applied to preserve the maximum value within local receptive fields and discard all other values. Gaussian dropout was introduced to prevent overfit and the feature maps were flattened to fully connected layers. Eventually, the softmax function was applied to achieve binary classification with a loss function of cross-entropy, leading to the identification of the images as either crystal or liquid. 

\section{Results}
An unsupervised machine learning method was applied to investigate the phase diagram of 2D complex plasmas. For a certain $\kappa$, we can arbitrarily assume a melting temperatures and divide the image sequences into samples of liquid and crystal accordingly. Obviously, if this assumed temperature approaches the lowest and highest temperature in the total samples, all the images are virtually regarded in one single thermodynamics status, resulting in a high accuracy in the validation. Except for these extreme scenarios, if the assumed melting temperature is different from the true melting temperature, the neural network is confused by the mixed samples and naturally results in a relatively low validation accuracy. However, if the assumed melting temperature coincides with the true melting temperature, the neural network can successfully recognize the different thermodynamics status of the 2D complex plasmas in the images based on the structures of the particle positions and thus results in a high validation accuracy~\cite{Nieuwenburg:2017}.

\begin{figure}[!ht]
	\includegraphics[width=18pc]{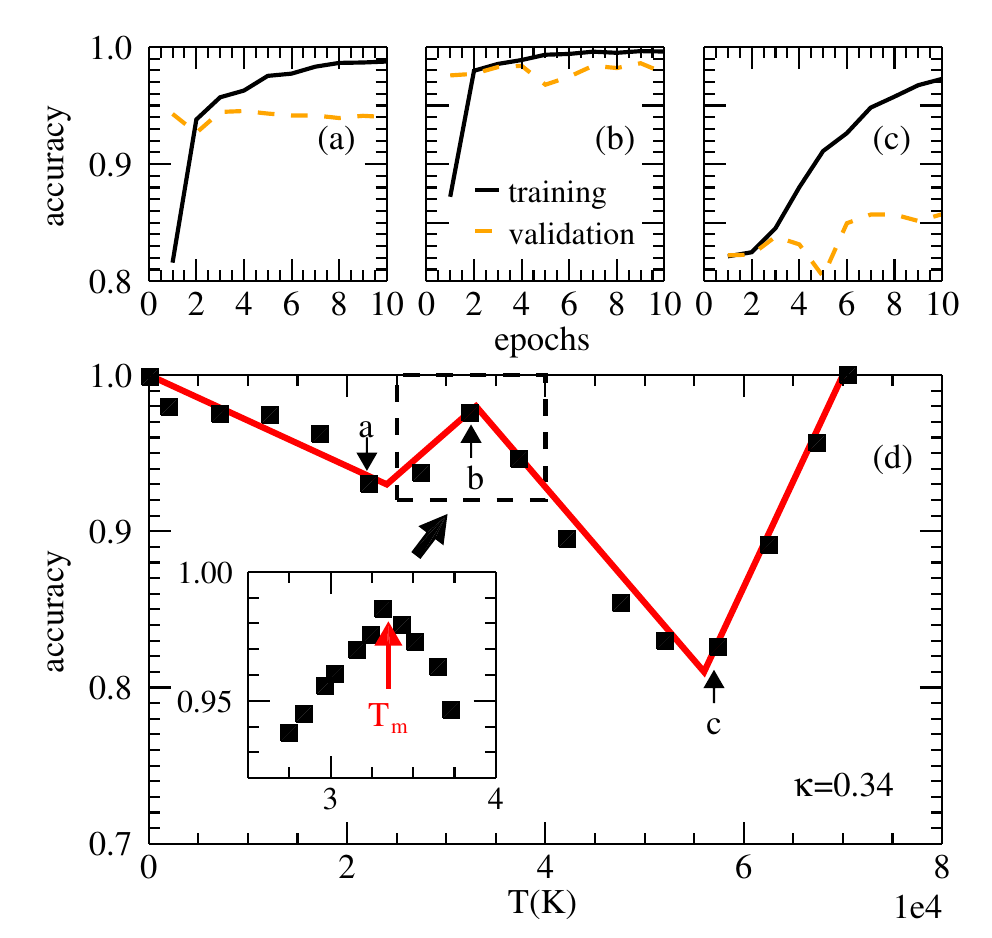}
	\caption{ Training and validation accuracy for the unsupervised machine learning method at $\kappa=0.34$. (a-c) The training and validation accuracy for the assumed melting temperature at $22240, 33170, 57390$~K. (d) The dependence of the validation accuracy on the assumed melting temperature, exhibiting a W-like shape. The red lines highlight the shape of this dependence. The peak of the validation accuracy is zoomed in the inset and the corresponding temperature is essentially the true melting temperature for the selected $\kappa$.}
	\label{fig:accuracy}
\end{figure}

The dependence of the validation accuracy on the assumed melting temperature is shown in Fig.~\ref{fig:accuracy} for $\kappa=0.34$. The images were divided into the training and validation samples by a ratio of $80\%$ to $20\%$. The evolution of the training and validation accuracy against epochs for three assumed melting temperature are shown in Fig.~\ref{fig:accuracy}(a-c). In the panel (a), the assumed melting temperature was lower than the true melting temperature, the training accuracy rose relatively fast and reached $0.99$ after training for $10$ epochs, while the validation accuracy was always lower than the training accuracy. As the assumed melting temperature coincided with the true melting temperature, shown in the panel (b), the training accuracy rose even faster than in panel (a) and reached unity after $5$ training epochs. More importantly, the validation accuracy was also close to unity for the whole training procedure. In the panel (c), the assumed temperature was much higher than the true melting temperature, the training accuracy rose relatively slowly and reached $0.98$ after $10$ epochs. However, the validation accuracy was significantly lower and fluctuated around $0.82$, indicating a wrong classification. 

The overall dependence of the validation accuracy on the assumed melting temperature exhibited a W-like shape, as shown in Fig.~\ref{fig:accuracy}(d). We zoomed the peak in the inset and refined the sampling step of the assumed melting temperature. The peak lay at the temperature of $33170$~K, which was the true melting temperature of the system for $\kappa=0.34$. Note that here we did not label the samples based on any features, which was usually required in the supervised machine learning methods.

\begin{figure}[!ht]
	\includegraphics[width=16pc]{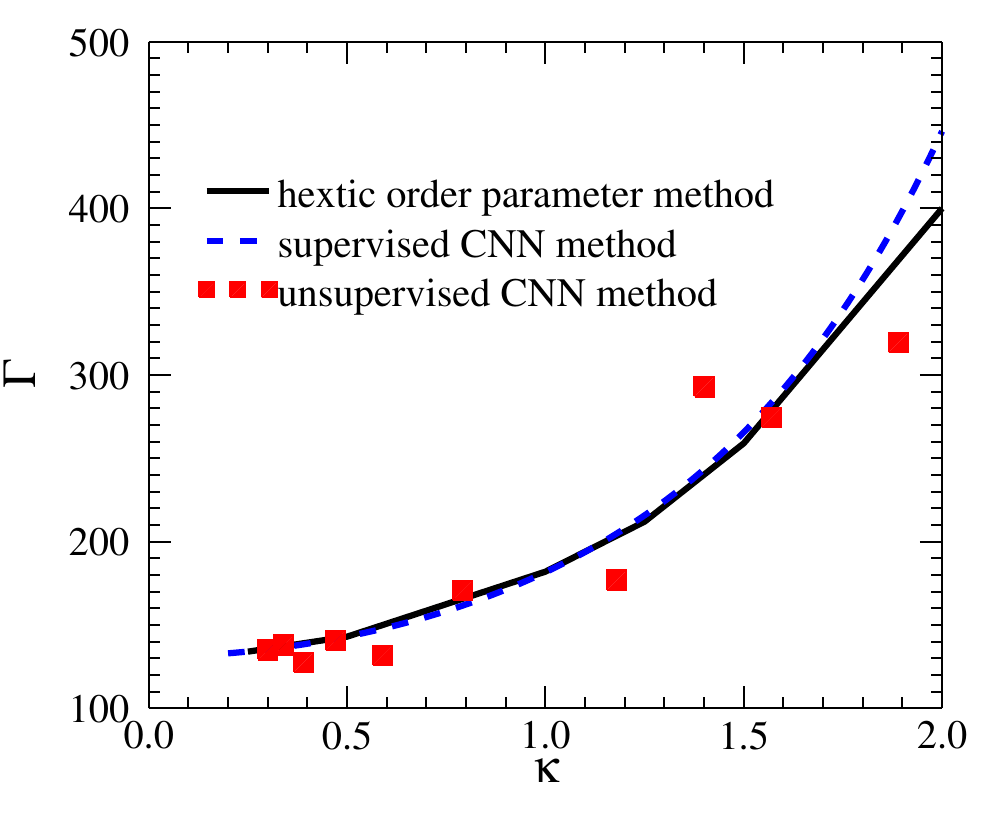}
	\caption{ Melting line in the phase diagram of the 2D complex plasma. The black line is obtained based on the hexatic order parameter~\cite{Hartmann:2005} and the dashed line is obtained by the supervised machine learning method~\cite{Huang:2022}. The red squares are the results by the unsupervised machine learning method reported in this paper.}
	\label{fig:phase}
\end{figure}

Finally, we repeated this procedure for different $\kappa$ and obtained the melting line in the phase diagram of the 2D complex plasmas. The results are shown as red squares in Fig.~\ref{fig:phase}. The solid line resulted from the analysis of the hexatic order parameter of the 2D complex plasma, where the melting temperature was defined such that $\overline{|\Psi_6|}=0.45$~\cite{Hartmann:2005,Schweigert:1999}.  The dashed line was the fitted curve to the melting temperature obtained by the supervised machine learning method, where the liquid and crystal were labeled before training at extreme temperatures~\cite{Huang:2022}. Our results generally agree with those obtained by the above two methods. The thermodynamic status in the 2D complex plasma can be identified by the transient particle structure alone.

\section{Conclusion and outlooks}
In this paper, we demonstrate that the unsupervised machine learning method can be applied to the identification of the melting line in the 2D complex plasmas. The resulting melting line coincides with those obtained from the analysis of hexatic order parameter and supervised machine learning method. It turns out that $\overline{|\Psi_6|}=0.45$ is indeed an suitable critical value of the order parameter to identify the melting transition of 2D complex plasmas.

As the samples do not need to be labeled based on any order parameter, such method can be potentially applied to study the phase diagram of the 3D complex plasmas, where the structures of the plasma crystal deviate from the standard bcc, fcc and hcp structure \cite{Zuzic:2006,Wang:2020}. Such deformed structures may be caused by the anisotropic effects such as ion drag or gravity, and thus can not be accurately constructed and labeled. The advantage of the unsupervised machine learning method shall overcome such obstacles. Besides, this method can also be applied to the experiment results as long as sufficiently large samples can be collected. We leave these for the future work. 

\section{Acknowledgments}
This work was supported by the National Natural Science Foundation of China (NSFC), Grant No.~$11975073$ \& $21035003$, and the Fundamental Research Funds for the Central Universities, Grant No. $2232023$G-$10$.


\end{document}